\documentclass[12pt]{article}
\usepackage{graphicx}
\usepackage{amsmath}
\usepackage{lipsum}
\usepackage{hyperref}
\usepackage{enumitem}
\usepackage{float}
\usepackage{enumitem}

\title{\textbf{Balancing Fixed Number of Nodes Among Multiple Fixed Clusters}}
\author{
 Paritosh Ranjan \\
  IBM  \\
  \texttt{paranjan@in.ibm.com} \\
  \and
 Surajit Majumder \\
  IBM  \\
  \texttt{surajit.majumder@ibm.com} \\
  \and
 Prodip Roy \\
  IBM  \\
  \texttt{prodipro@in.ibm.com} \\
    \and
 Bhuban Padhan \\
  IBM  \\
  \texttt{bhubanpadhan@in.ibm.com} \\
}
\date{\today}

\begin{document}

\maketitle

\begin{abstract}
Cloud infrastructure users often allocate a fixed number of nodes to individual container clusters (e.g., Kubernetes, OpenShift), resulting in underutilization of computing resources due to asynchronous and variable workload peaks across clusters. This research proposes a novel system and method for dynamic rebalancing of a fixed total number of nodes among multiple fixed clusters based on real-time resource utilization thresholds. By introducing a Node Balancing Cluster Group (NBCG), clusters are grouped and allowed to dynamically share nodes through a controlled reallocation mechanism, managed by a Node Balancing Cluster Balancer and a Resizing Rule Engine. The system identifies overutilized and underutilized clusters using threshold parameters, and reassigns nodes without incurring additional provisioning costs. If reallocation causes a violation of utilization thresholds, the system reverses the operation to maintain cluster stability. The proposed architecture not only optimizes resource utilization and operational cost but also introduces a strategic advantage for cloud service providers like IBM Cloud. Unlike existing solutions, this approach enables intra-account node sharing across clusters with strict adherence to user-defined constraints and ensures consistent cluster state management. This invention has the potential to significantly reduce computing resource waste and position IBM Cloud services as more efficient and competitive.

\end{abstract}

\section{Introduction}

Cloud computing has become the de facto application and infrastructure management platform. Organizations and users create accounts in cloud platforms such as IBM Cloud, Google Cloud, AWS etc. These accounts are then used to provision resources as Software as a service, Platform as a service, Infrastructure as a service etc. The provisioned cloud resources are then used by the users and the organizations to run and operate their software.
The cloud providers provide clusters for hosting containers e.g., Kubernetes, Openshift etc. These clusters are group of nodes on which the containers run.
Usually, each cluster has a fixed number of worker nodes assigned to it. The cloud user or cloud account can have multiple clusters and would assign fixed number of nodes to each cluster. Different clusters can have different timings of peak usage. So, in case of any sudden increase in demand by any of the clusters in this group, other computing resources from under-utilized clusters can be utilized, thus saving cost and unnecessary provisioning of more computing resources.
If any cloud account or user has several clusters with fixed number of nodes in each cluster then the free computing resource in each cluster would add up to significant amount of computing resources at any point of time.
This invention aims to reduce this wastage of computing resources and money by optimizing the usage of computing resources among multiple clusters.

\section{Brief Description of the Invention}

This invention introduces a method and system for dynamically balancing a fixed number of computing nodes among a fixed number of container-hosting clusters (e.g., Kubernetes or OpenShift) within a cloud environment. A logical grouping mechanism, called the Node Balancing Cluster Group (NBCG), allows multiple clusters to participate in node sharing based on real-time workload demands.

Each cluster initially receives a fixed number of nodes. When added to an NBCG, node allocation is no longer static but governed by a centralized component called the Node Balancing Cluster Balancer. This balancer continuously monitors cluster resource utilization using a Node Balancing Cluster Locator and makes reallocation decisions based on thresholds defined by the user via a Resizing Rule Engine.

When a cluster exceeds its upper resource threshold (THigh), the system identifies another cluster with usage below the lower threshold (TLow). The node from the underutilized cluster (CLow) is drained of workloads using the Node Retriever, deprovisioned, and reassigned to the overutilized cluster (CHigh)—provided this does not push CLow above THigh after node removal. If no suitable cluster is found, the rebalancing is aborted, and the system restores the original state.

Additional features include:
\begin{enumerate}
    \item Adding or removing clusters from the NBCG at any time,
    \item Tracking the origin of reassigned nodes,
    \item Reinstating a cluster’s original node configuration when removed from the NBCG.
\end{enumerate}

\section{Reduction to Practice}

This invention describes a method to balance a fixed number of clusters (hosting containers) with a fixed number of nodes balanced between the clusters.

\textbf{Steps}
\begin{enumerate}
    \item A node balancing cluster group would be defined, and clusters would be assigned to the node balancing group. The clusters which would be part of the node balancing cluster group would be able to balance nodes among themselves as needed.
    \item The user would be able to create multiple node balancing cluster groups and assign clusters to them. One cluster could be assigned to only one node balancing cluster group.
    \item Each cluster would originally be provisioned with a fixed number of nodes. However, once the cluster becomes part of the node balancing cluster group, the number of nodes provisioned in the cluster would be decided and controlled by the node balancing cluster balancer. 
    \item Any cluster can be moved out of the node balancing cluster group at any time to stop its participation in node balancing cluster group by node balancing cluster node retriever. The cluster can be added again to the node balancing cluster group if needed.
    \item The node balancing cluster balancer would take decisions of deprovisioning or provisioning nodes using node balancing cluster resizing rule engine and based on thresholds provider by the user.
    \begin{enumerate}
        \item If the total resource utilization threshold of any cluster CHigh would be greater than THigh, then node balancing cluster resizing rule engine will search the clusters whose utilization is below TLow via node balancing cluster locator and pick the cluster CLow with lowest utilization among them. 
    \end{enumerate}
    \item Node balancing cluster node retriever will evict pods from the lowest utilized node of this CLow cluster, deprovision this node, and then calculate CLow cluster’s resource utilization via node balancing cluster resource utilization calculator. 
    \begin{enumerate}
        \item If the resource utilization of the low utilization CLow cluster after node deprovisioning would remain below THigh even after deprovisioning the node, then one node would be provisioned in the first cluster CHigh whose THigh was greater than threshold.
        \item However, if the resource utilization of CLow cluster after node deprovisioning would be high than THigh, then this node balancing would not be permitted, and the node would be reprovisioned to the CLow cluster. This check would be then performed for all other clusters having resource utilization lower than TLow, until the node balancing is possible or all clusters are checked.
    \end{enumerate}
    \item So, the nodes of the clusters would get rebalanced without any extra cost of provisioning more computing resources.
\end{enumerate}

\begin{figure}
    \centering
    \includegraphics[width=0.9\linewidth]{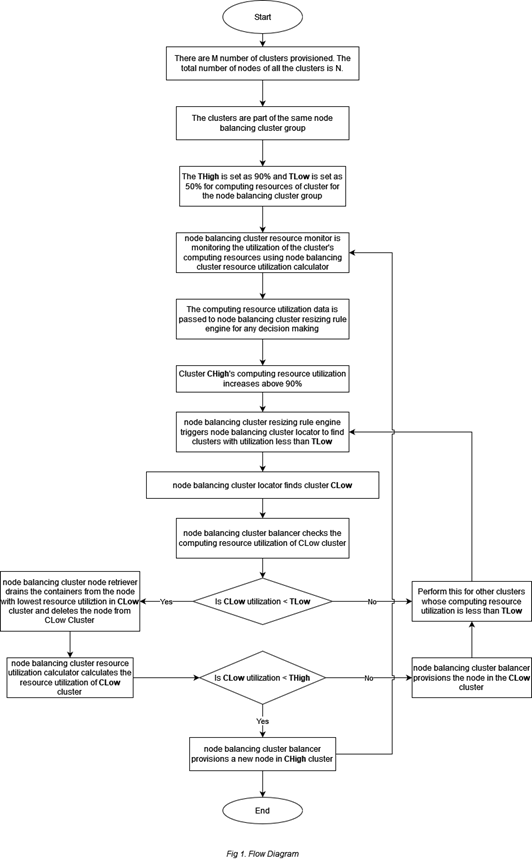}
    \caption{Flow-Diagram}
    \label{fig:{Flow-Diagram}}
\end{figure}

To demonstrate the feasibility and effectiveness of the proposed invention, a prototype was implemented within a simulated cloud environment using Kubernetes clusters. The system comprised the following key components, developed as microservices and deployed on a control plane:

\begin{figure}
    \centering
    \includegraphics[width=0.9\linewidth]{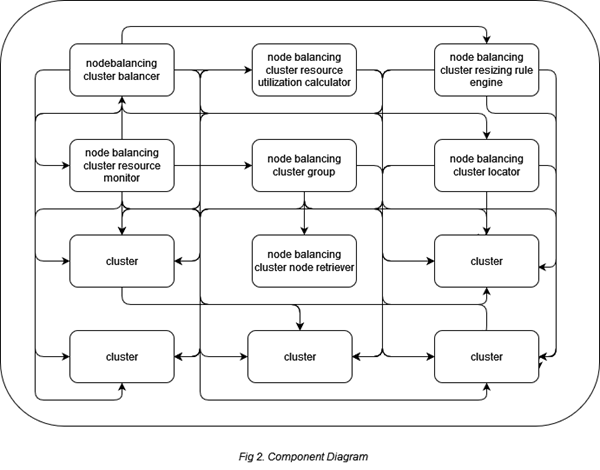}
    \caption{Component-Diagram}
    \label{fig:Component-Diagram}
\end{figure}

\begin{enumerate}
    \item Node Balancing Cluster Group Manager – Enabled the grouping of clusters into logical balancing units and allowed dynamic addition and removal of clusters from the group.
    \item Node Balancing Cluster Balancer – A centralized service that continuously monitored resource utilization metrics (CPU, memory) collected via Kubernetes Metrics Server and Prometheus.
    \item Node Balancing Cluster Locator – Implemented to identify clusters whose utilization exceeded the upper threshold (THigh) and those below the lower threshold (TLow).
    \item Resizing Rule Engine – Configured with user-defined thresholds and policies. It evaluated conditions to determine whether node reallocation could proceed without destabilizing the source cluster.
    \item Node Retriever – Used Kubernetes eviction APIs and node taints/tolerations to safely drain pods from the least utilized node in the donor cluster, followed by node deprovisioning using the cloud provider’s API.
    \item Node Provisioner – Managed the provisioning of the retrieved node into the recipient cluster, ensuring compatibility and updating internal tracking systems for node origin.
\end{enumerate}

\section{Advantages of the Invention}

\begin{enumerate}
    \item Optimized Resource Utilization : The invention enables dynamic reallocation of underutilized nodes across clusters, significantly improving overall resource usage without requiring additional infrastructure.
    \item Cost Efficiency : By reducing the need to overprovision computing resources during peak load times, the system minimizes operational costs and maximizes the ROI on existing infrastructure.
    \item Scalability Without Overhead : The balancing system allows clusters to handle peak demand by borrowing nodes from others, eliminating the need for expensive and time-consuming auto-scaling operations.
    \item Non-Disruptive Operation : Nodes are drained and migrated using Kubernetes-native mechanisms, ensuring zero downtime for running applications and preserving service continuity during rebalancing.
    \item Policy-Driven Control : Administrators retain control through user-defined thresholds and policies, enabling fine-grained tuning of balancing behavior based on workload characteristics.
    \item Cluster Autonomy and Reversibility : Each cluster can be added to or removed from the balancing group at any time. Upon removal, the system restores the original node configuration, ensuring flexibility and independence.
    \item Vendor Differentiation : This feature offers a competitive edge for cloud service providers like IBM Cloud, enhancing their Kubernetes or OpenShift offerings with intelligent, cost-saving infrastructure management.
\end{enumerate}

\section{Conclusion}
This invention presents a novel and practical approach to balancing a fixed number of nodes among a fixed number of container clusters within a cloud environment. By enabling intelligent and dynamic reallocation of nodes based on real-time resource utilization thresholds, the system maximizes efficiency, reduces costs, and eliminates the inefficiencies associated with static resource provisioning.

The architecture is both flexible and robust—allowing seamless entry and exit of clusters from the balancing group, maintaining original cluster states, and ensuring consistent performance. The reduction to practice further validates the system’s real-world applicability and potential for integration into commercial cloud platforms.

As organizations increasingly rely on multi-cluster cloud environments, this invention addresses a critical gap in cluster resource management, providing a scalable and cost-effective solution that aligns with the operational and economic goals of modern cloud infrastructure providers and users.

\section{Acknowledgment}

We would like to express our sincere gratitude to all individuals and organizations who have contributed to the success of this research. We acknowledge the invaluable support from the IBM team, whose resources and expertise have greatly enhanced this project.
Special thanks to Prodip Roy (Program Manager IBM) for their insightful feedback, guidance, and encouragement throughout the development of this work.

\section{References}
\renewcommand\refname{}

\end{document}